# COSMIC IRONY: SETI OPTIMISM FROM CATASTROPHES?

Milan M. Ćirković
Astronomical Observatory Belgrade
Volgina 7, 11160 Belgrade
SERBIA AND MONTENEGRO
arioch@eunet.yu

**ABSTRACT.** Classical arguments for skepticism regarding the Search for ExtraTerrestrial Intelligence (SETI) are critically examined. It is suggested that the emerging class of "phase transition" astrobiological models can simultaneously account for all available astrophysical and biological evidence, explain several unresolved puzzle in Earth sciences, and rationally justify current and future SETI projects. In particular, the hypothesis of Annis that local gamma-ray bursts drive the astrobiological phase transition deserves to be further quantitatively elucidated. Some epistemological and ethical ramifications of such a model are briefly discussed.

**KEYWORDS:** Astrobiology, anthropic principle, gamma-ray bursts, philosophy of astronomy

*It is the thunderbolt that steers the course of all things.*

Heraclitus of Ephesus, cca. 550 BC

## INTRODUCTION: KEY ANTI-SETI ARGUMENTS

It is hard to deny that the Search for Extra-Terrestrial Intelligence (SETI) is one of the major scientific adventures in the history of humankind. However, four decades of serious SETI projects have not given results, in spite of the prevailing "contact optimism" of 1960s and early 1970s, motivated largely by uncritical acceptance of the Drake equation. Conventional estimates of that period spoke of about $10^6$ - $10^8$ (!) advanced societies in the Milky Way forming the "Galactic Club" (Bracewell 1975). Nowadays, even the greatest SETI optimists have abandoned such fanciful numbers, and settled to a view that advanced extraterrestrial societies are rarer than previously thought. Today, it is widely recognized that the "contact pessimists" have a rather strong position (Hanson 1999; Bostrom 2002); most of recent scholarly monographs on the subject are strongly sceptical towards the possibility of finding complex intelligent life elsewhere (e.g. Ward and Brownlee 2000; Webb 2002). Why is that so?

Anti-SETI "rollback" of the late 1970s and 1980s has been essentially based on two widely discussed arguments:

(1) Tsiolkovsky-Fermi-Viewing-Hart-Tipler's question[1] "Where are they?" in its modern von-Neumann-probe rendering, and
(2) Carter's (1983) "anthropic" argument.

---

[1] Stephen Webb, in his recent monograph, so far the best historical introduction into the "Great Silence" problem (Webb 2002), dubbes the relevant question Tsiolkovsky-Fermi-Viewing-Hart's. We find it only just to add Tipler to list, since his von Neumann probe setup gives the whole problem completely new flavor (Tipler 1980). Of course, it is best known simply as "Fermi's paradox" (which we shall use, for the sake of brevity, in the rest of the paper).

Tsiolkovsky, Fermi, Hart, and their supporters argue from the absence of extraterrestrials on Earth and in the Solar System, and the fact that they have, *ceteris paribus*, more than enough time in the history of the Galaxy to visit, either in person or through their self-replicating probes. Characteristic time for colonization of the Galaxy, according to these investigators is $10^6 - 10^8$ years, making the fact that the Solar System is (obviously) not colonized hard to explain, if not for the absence of extraterrestrial cultures. On the other hand, Carter's "anthropic" argument ("argument from ignorance" would be a better label here) tries to infer conclusions from the possible relationships between the alleged astrophysical ($\tau_*$) and biological ($\tau_l$) timescales. In the Solar system, $\tau_* \approx \tau_l$, within the factor of two. However, in general, it should be either $\tau_l \gg \tau_*$ or $\tau_* \gg \tau_l$ for two uncorrelated numbers. In the latter case, however, it is difficult to understand why the very first inhabited planetary system (that is, the Solar System) exhibits $\tau_* \approx \tau_l$ behaviour, since we would then expect that life (and intelligence) arose on Earth, and probably at other places in the Solar System, much earlier than they in fact did. This gives us probabilistic reason to believe that $\tau_l \gg \tau_*$ (in which case the anthropic selection effects explain very well why we do perceive the $\tau_* \approx \tau_l$ case in the Solar System). Thus, according to Carter, the extraterrestrial life and intelligence have to be very rare, which is the reason why we have not observed them so far.

It seems clear that any general astrobiological scenario capable of dissolving these two crucial anti-SETI arguments would not only undermine the position of "contact pessimists", but also offer a valuable methodological guidelines for the SETI projects themselves. With this in mind, in the rest of this paper we suggest a "catastrophic" astrobiological scenario based on the phase-transition of Annis (1999), which is capable of both refuting the anti-SETI arguments and give us significant rational basis for optimism in our searches for intelligent life elsewhere in the Milky Way.

**EVOLUTIONARY APPROACH TO THE ASTROBIOLOGICAL STATUS OF OUR GALAXY**

First, it is necessary to mention that both Fermi's "paradox" and Carter's argument have been criticized so far. Thus, for instance, the colonization timescale is still largely uncertain; for instance, diffusion models of Newman and Sagan (1981) give the relevant timescale as $\sim 10^9$ years, which would correspond to the naive answer one is expected to give on the Fermi's question: *They are still on the way*! Second, the issue of motivation of colonizers, and particularly their von Neumann probes is much less clear and unambiguous than the "contact pessimists" would have us believe. Notably, as suggested by Brin (1983) in his seminal review, the "deadly probes" scenario (the idea that the dominant behavior of self-replicating probes is destruction of nascent civilizations, not colonizing) is one of just a few theoretically satisfactory explanations of the "Great Silence". In a similar vein, Kinouchi (2001) has recently argued that the phenomenon of persistence, well-known from statistical physics, holds the key for explanation of the apparent absence of extraterrestrial civilizations; in this picture, Galactic colonization by advanced ETIs could have already last for quite some time without influencing the Solar System. Wilson (1994) has persuasively criticized Carter's usage of the anthropic principle to show that life is rare in the universe.

However, the most important line of thought which can easily defeat both Fermi's and Carter's arguments lies in investigation of hidden temporal assumptions in these arguments. Fermi et al. suppose that the history of the Galaxy is uniformitarian, in the sense that advanced technological communities could arise at any point in the Galactic history (except, perhaps, the first billion years or so, when the metallicity was too low; see Lineweaver 2001). Similarly, Carter assumes that the only relevant astrophysical timescale is the Main Sequence stellar lifetime. Uniformitarianism has not shown brightly in astrophysics and cosmology, at least since the demise of the classical steady-state theory in mid-1960s (Kragh 1996). Today we are quite certain that evolutionary properties of astrophysical systems are from time to time guided by processes either rare or unique (like the primordial nucleosynthesis or the reionization of intergalactic medium) or occuring at timescales so much vaster than the timescales of human civilization that the probability of actually observing them is nil (like the recently computed evolution of M-dwarf stars[2]). In the specific case, if the phase-transition model sketched in a brilliant short paper of Annis (1999; see also Clarke 1981) is correct—as we have more and more reasons to believe—the relevant timescale is the one describing intervals between the major Galactic-wide catastrophes, **precluding** the complexification of planetary biospheres and, consequently, the development of intelligent observers. There are several plausible candidates for this **global regulation mechanism**. The strongest, as suggested by Annis in his ingenuous study, are gamma-ray bursts (henceforth GRBs), which accompany either a coalescence of binary neutron stars or explosions of super-massive stars, also known as the *hypernovae* (for a review of GRB mechanisms, see Piran 2000).

Astrobiological effects of GRBs have been investigated recently in a number of papers (Thorsett 1995; Dar 1997; Scalo and Wheeler 2002), and much of the older literature dealing with effects of supernova explosions is useful in this case too (after appropriate scaling, of course; see, for instance, Tucker and Terry 1968; Ruderman 1974; Clark, McCrea, and Stephenson 1977). It seems that each GRB is surrounded by a "lethality zone" in which its effects are deadly for complex lifeforms (eucaryotes); according to Scalo and Wheeler (2002). the radius of this zone is ~14 kpc, rather large in comparison to the Galactic habitable zone. The exact effects of a GRB within a "lethality zone" are still somewhat controversial, but it is clear that there will be at least two deadly effects capable of causing mass extinctions: **(A)** creation of nitrogen-oxides (usually denoted by $NO_x$) in the upper atmosphere, which will destroy the ozone layer for thousands of years, thus enormously increasing UV radiation at planetary surface; and **(B)** creation of a longer delayed pulse of cosmic rays, which penetrate the atmosphere (and even rocks and soil up to several km of depth) and cause various sorts of damage to biological materials. Both these effects are prolonged in comparison to the GRB itself, thus affecting not only the hemisphere directed toward the source. In fact, the consequences in biological domain may last many generations, especially when one considers such effects as increase in frequency of cancers, and occasionally very long interval needed for a species to die out when its population decreases below the so-called minimum viable population. Very recently, it has been suggested that one of the canonical "big five" mass extinctions in Earth's history, namely the late-Ordovician one, was caused by a local gamma-ray burst (Melott et al. 2003).

Other suggested regulation mechanisms are the climatic change due to interaction with Galactic spiral arms (Shaviv 2002), neutrino-induced extinctions (Collar 1996), or Galactic tides leading to the Oort comet cloud perturbations (e.g. Clube and Napier 1990; Rampino 1998).[3] Their common property is that they are **global**, i.e. influencing the entire Galactic habitable zone, or a large portion of it. Moreover, they **reinforce** each other; in other words, the total risk function is a sum of risk functions for each specific threat to biological systems. If the further research confirm that GRB-regulation however, has another desirable property: quantifiable **secular evolution**, which explains **our existence** at this particular epoch of the Galactic history. Notably, cosmology suggests the rate of GRBs behaves, on the average, as $\propto \exp(-t/\tau)$, with the time-constant $\tau$ of the order of $10^9$ yrs (Annis 1999). Namely, as noticed by Norris (2000), we have to ensure that there is no "overkill" as far as the regulation mechanisms are concerned, and that our own existence is explicable in naturalistic terms. This is readily achieved within the framework of the GRB-dominated phase-transition picture: cosmology assures us that the average rate of GRBs increases with redshift, i.e. decreases with cosmic time. When the rate of catastrophic events is high, there is a sort of quasi-equilibrium state between the natural tendency of life to spread and complexify and the rate of destruction and extinctions governed by the regulation mechanism(s). When the rate becomes lower than some threshold value, intelligent and space-faring species can arise in the interval between the two GRB-induced extinctions, and the Galaxy experiences a phase transition: from essentially dead place, with pockets of low-complexity life restricted to planetary surfaces, it will, on a very short Fermi-Hart-Tipler timescale, become filled with high-complexity life. We are living within that interval of exciting time, in the state of **disequilibrium** (Almár 1992), on the verge of the Galactic phase transition.[4] In Heraclitus' aphorism quoted above, Galactic GRBs are the "thunderbolt" which steers the course of all things astrobiological.

---

[2] For this fascinating subject in theoretical astrophysics, see Laughlin, Bodenheimer, and Adams (1997).
[3] The idea of Clarke (1981) that nuclear outbursts—similar to the ones observed in Seyfert galaxies—from the core of the Milky Way can lead to devastation of habitable planets throughout the Galaxy has been, historically, the first global-regulation mechanism proposed. However, it seems to be abandoned as we learn more about the center of our Galaxy. (For variations—now of "only" historical importance—on the same theme see Clube 1978; LaViolette 1987.)

It is clear that this class of models effectively removes the threat to ETIs from both Fermi-Hart-Tipler and Carter's arguments. Elsewhere in the Galaxy there are other planets with the level of complexity achieved more or less similar to the terrestrial one. There simply was not enough time for them to come to us, since the astrobiological history—as far as complex metazoans are concerned—is different and significantly shorter from the history of dark matter, stars, and gas clouds which constitute the physical structure of the Galaxy. Local astrobiological clocks can tick at various rates, but they are all from time to time reset by the global regulation mechanism(s). But Fermi's question is **rapidly becoming pertinent**, when we realize that during the phase transition many advanced intelligent societies are bound to develop, but they are not all bound to expand to their utmost limits (that is, to colonize the Galaxy) within the same interval of time.

On the other hand, the very existence of well-defined astrophysical and biological timescales is an unwarranted assumption of Carter's argument. This assumption is wrong in the context of the phase-transition models. The real timescales are specific to each planetary system, depending on such factors as the location of the system in the Galactic habitable zone (GRB distribution having a spatial, as well as temporal aspect!), peculiarities of the local environment (notably the density and distribution of cometary and asteroidal material presenting the impact hazard), and—of crucial importance—the epoch of Galactic history. In other words, there is no physical reason why on planet A, at galactocentric distance $R_A$ and at epoch $t_A$ we could not have $\tau_l \gg \tau_*$ while on planet B (characterized by $R_B$, $t_B$, and probably some other astrobiological parameters) we could have $\tau_l \ll \tau_*$. The dependence on the epoch is particularly important; to paraphrase the title of the controversial book by Ward and Brownlee (2000), **Earths might be rare in time, not in space**. This sort of models can also shed some new light on the Drake equation (Walker and Ćirković 2003; Ćirković 2003).

**LESSONS: ASTROPHYSICS OF OPEN SYSTEMS**

The phase-transition models represent a natural extension of the contemporary pictures developed in Earth sciences. We have recently learned that impacts of extraterrestrial bodies had tremendous influence upon the evolution of the biosphere (Raup 1991, 1999; Clube and Napier 1990). Even the mundane phenomena like cloudy skies are influenced by such previously un-dreamt of factors like the low-energy cosmic ray flux (e.g. Carslaw et al. 2002). In general, this is a part of the wider culturological tendency to relinquish the notion of a closed-box system in favor of **open (complex) systems**.

Even in the connection with such rather well-understood threats to life on Earth as collisions with comets or asteroids, a wider connection is sought in investigation possible influences from without the Solar System (Matese and Whitman 1992; Matese and Whitmire 1996; Rampino 1998). For instance, it has been investigated whether the giant molecular clouds, concentrated toward the Galactic plane, or even the tidal force of the Galactic disk itself can produce sufficient perturbations of the Oort cloud in order to cause impacting comet showers. Crossing of Galactic spiral arms has been repeatedly argued to be dangerous from the planetological, climatological, and biological points of view (e.g. Marochnik 1983; Shaviv 2002). There is no reason to think that this tendency will be arrested or rejected in the years to come, as both astrophysicists and planetary scientists on one side and biologists and ecologist on the other, probe more and more subtle and minuscule effects, and their computer simulations become richer and richer with seemingly disconnected phenomena. Even the classical Gaia hypothesis (e.g., Lovelock 1988) is rather moderate in this respect, when astrobiological issues are concerned. We envisage a tight interconnections of (astro)physical and (astro)biological aspects of reality on scales as large as the Milky Way galaxy. The closed-system innocence (or "splendid isolation") of Earth and its biosphere is forever lost.

---

[4] Notice that the anthropic selection effect (cf. Bostrom 2002) readily explains why that is so, in spite of the very low *a priori* probability. Humans could not arise prior to the phase transition, since there was no time for high-complexity life to evolve without being destroyed by cosmic rays and other detrimental consequences of GRB regulation (or cumulative effects of impacts, close SNe, spiral-arms crossings, and other calamities). On the other hand, we could not arise later from the phase transition epoch for the same reason one does not expect to find a previously unknown stone-age tribe in the present-day Europe: high-complexity ecological niches do not allow spontaneous emergence of new lower-complexity lifeforms.

Two important practical conclusions for SETI projects are the following. (1) Phase-transition models suggest that extraterrestrial civilizations are vastly more likely to be of appropriate age for communication than could be inferred by naive application of the Drake equation. That is, we do not need to worry further about the sensibility of attempting to communicate with beings billion or more years older from us. Phase-transition models predict that all Galactic civilizations will be of similar age, i.e. the time elapsed since the last "resetting" of the astrobiological clock. (2) Future detailed numerical phase-transition theory (Ćirković and Sandberg, in preparation) will enable assigning statistical weights to various SETI target stars and planetary systems depending on their ages and orbital elements in the Galaxy. For instance, it is intuitively clear that we are more likely to encounter advanced biospheres in the outer parts of the Galactic Habitable Zone than in the inner parts, since the density of GRBs and supernovae declines sharply with the galactocentric radius, so outer regions are more likely to stay undisturbed for sufficiently long time to allow for biogenesis and noogenesis. However, additional effects of metallicity, spiral-arm crossings, etc. have to be taken into account in the future numerical work in this direction.

**CONCLUSIONS**

The phase-transition models offer a hope of reconciling both our astrophysical knowledge and negative SETI results on one hand, with naturalistic explanations for biogenesis and noogenesis and the Copernican Principle on the other. Thus, Fermi's paradox is explained as a part of much wider astrobiological paradigm, which is certainly desirable from the methodological point of view.

The price paid is obviously high, but not in usual epistemical sense, but rather in the ethical one. It means that enormous destruction of life is taking place in what is conventionally portrayed as peaceful and hospitable universe. GRBs occur once per day on the average, and they sample practically all galaxies within our particle horizon. Thus, if the picture suggested by the phase-transition model is correct, we are witnessing immense destructions of complex lifeforms (some of them undoubtedly intelligent, but not reaching the "immunity level") throughout the universe on a daily basis! This cannot fail to be a depressing thought. Out past light-cone is full of slaughterhouses on an unprecedented level. Statistically, an absolutely staggering majority of all lifeforms ever emerging are exterminated by random and violent catastrophes of astrophysical origin. The universe might be a significantly more cruel and inhospitable environment than it's usually assumed. It is quite possible that simple organisms analogs to *Deinococcus Radiodurans* (Battista 1997) are by far the most prevalent form of life in the universe on the average.

But on the balance, it is still hard to find this scenario worse than any naturalistic alternative. As noted by many biologists, the role of mass extinctions in the history of life here on Earth was ambiguous, both destructive and constructive (e.g., Raup 1994); we would not have been here almost certainly if comet or asteroid did not struck Earth 65 Myr ago causing, among other things, the end of the epoch of giant reptilians. It is far from certain that, had this event not happened, the intelligence would have developed on Earth (by now, at least).

This may mean, on the other hand, that those who survive are even more valuable than conventionally thought. On the background of natural disasters, even a single anthropogenic disaster (like the looming ecological dangers, nuclear warheads, or nanotechnology) is simply the one disaster too many. If properly understood, the astrobiological models can teach us the rarity of **values** – the ultimate products of generic creative and intelligent minds. In futures, these values will almost certainly (by one or another roughly human-comparable civilization) be spread to all corners of the Galaxy, and possibly even beyond. It is these values that we ought to seek in our peers through the SETI programs, but also – and foremostly – to advance in ourselves through the entire scientific and artistic endevor of humanity.

**Acknowledgements.** The author thanks Larry Klaes, Robert J. Bradbury, Ivana Dragićević, Nick Bostrom, and Richard Catcart for useful and friendly discussions. Technical help of Vjera Miović, Ivana Dragićević, Branislav Nikolić, Ivan Almár, Vesna Milošević-Zdjelar, Saša Nedeljković, Srdjan Samurović, and Mark A. Walker is also kindly appreciated.

8